\documentclass{article}

\usepackage{citesort,color,graphicx,epsfig,pifont,xspace,amsfonts,amssymb,curves,url,mathrsfs}

\newtheorem{assum}{Assumption}

\def\cost(#1){\mathify{\| #1 \|}}

\newtheorem{ex}{Example}

\newtheorem{problem}{Problem}

\newcommand{\EG}{{\em e.g.}\xspace}
\newcommand{\EA}{{\em et al.}\xspace}

\newcommand{\comment}[1]{}

\newcommand{\binom}[2] 
	{\left(\begin{array}{c}#1 \\ #2\end{array}\right)}

\newlength{\bletn}

\title{Reverse Engineering of Molecular Networks from a Common Combinatorial Approach}

\author{Bhaskar DasGupta\\
Department of Computer Science \\
University of Illinois at Chicago \\
Chicago, IL 60607 \\
Email: {\tt dasgupta@cs.uic.edu} \\
\and
Paola Vera-Licona \\
Institut Curie, Paris Cedex 05, France\\
Email: {\tt paola.vera-licona@curie.fr} \\
\and 
Eduardo Sontag \\
Mathematics Department \\
Rutgers University \\
Piscataway, NJ 08854\\
Email: {\tt sontag@math.rutgers.edu} 
}

\begin{document}

\maketitle

\section{Introduction}

The understanding of molecular cell biology requires insight into the
structure and dynamics of networks that are made up of thousands of interacting
molecules of DNA, RNA, proteins, metabolites, and other components.
One of the central goals of systems biology is the unraveling of the as yet
poorly characterized complex web of interactions among these components.
This work is made harder by the fact that new species and interactions are
continuously discovered in experimental work, necessitating the development of
adaptive and fast algorithms for network construction and updating.
Thus, the ``reverse-engineering'' of networks from data has emerged as one of
the central concern of systems biology research.

A variety of reverse-engineering methods have been developed, based on tools
from statistics, machine learning, and other mathematical domains.
In order to effectively use these methods, it is essential to develop an
understanding of the fundamental characteristics of these algorithms.
With that in mind, this chapter is dedicated to the reverse-engineering of
biological systems.

Specifically, we focus our attention on a particular class of methods for
reverse-engineering, namely those that rely algorithmically upon the so-called
``hitting-set" problem, which is a classical combinatorial and computer
science problem, Each of these methods utilizes a different algorithm in order
to obtain an exact or an approximate solution of the hitting set problem.  We
will explore the ultimate impact that the alternative algorithms have on the
inference of published \textit{in silico} biological networks.

\section{Reverse Engineering of Biological Networks}
\label{DifMethods}

Systems biology aims at a systems-level understanding of biology, viewing
organisms as integrated and interacting networks of genes, proteins, and other
molecular species through biochemical reactions that result in particular form
and function (phenotype).  Under this ``system'' conceptualization, it is the
interactions among components that gives rise to emerging properties.

Systems-level ideas have been a recurrent theme in biology for several
decades, as exemplified by Cannon's work on homeostasis~\cite{Cannon},
Wiener's biological cybernetics~\cite{Wiener}, and Ludwig von Bertalanffy's
foundations of general systems theory~\cite{Bertalanffy}.  So what has brought
systems biology to the mainstream of biological science research in recent
years?  The answer can be found in large part in enabling technological
advances, ranging from high-throughput biotechnology (gene expression arrays,
mass spectrometers, etc.) to advances in information technology, that have
revolutionized the way that biological knowledge is stored, retrieved and
processed.
 
\begin{figure}[htbp]
\begin{center}
\includegraphics[width=4 in]{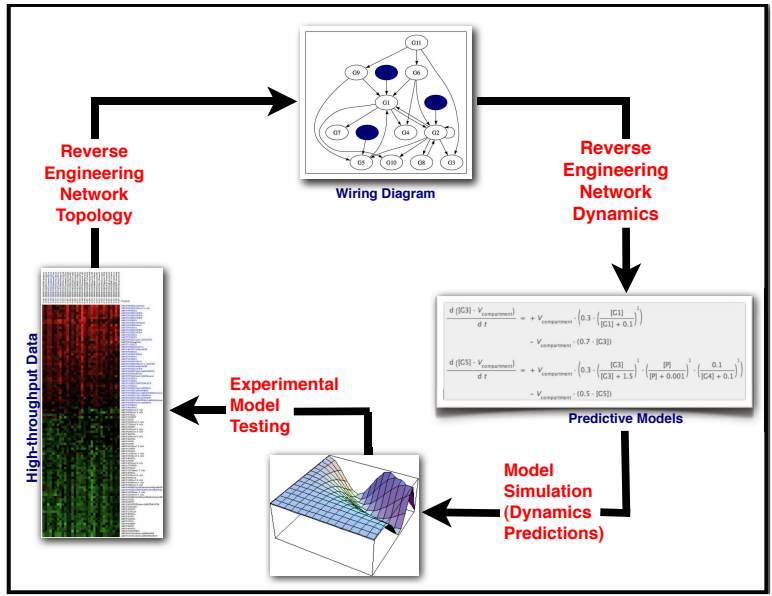}
\caption{Iterative Process in Systems Biology.} 
\label{SysBio}
\end{center}
\end{figure}

A systems approach to understanding biology can be described as an iterative
process which includes: (1) data collection and integration of all available
information (ideally, regarding all the components and their relationships in
the organism of interest), (2) system modeling, (3) experimentation at a
global level, and (4) generation of new hypotheses (see Fig.~\ref{SysBio}).  

The current chapter focuses on the system modeling aspects, and, specifically,
on the top-down modeling approach broadly known as the biological
``reverse-engineering'', which can be very broadly described as follows:

\begin{quote}
\emph{The biological reverse engineering problem is that of analyzing a given
  system in order to identify, from biological data, the components of the
  system and their relationships.}
\end{quote}

In broad terms, there are two very different levels of representation for
biological networks. They are described as follows.

\vspace*{0.1in}
\noindent\underline{\bf (a) Network Topology Representations}

\vspace*{0.1in}
Also known as ``wiring diagrams'' or ``static graphs'', these are coarse diagrams
or maps that represent the connections (physical, chemical, or statistical)
among the various molecular components of a network.  At this level, no
detailed kinetic information is included.  A network of molecular
interactions can be viewed as a graph: cellular components are nodes in a network,
and the interactions (binding, activation, inhibition, etc.) between these
components are the edges that connect the nodes.  A reconstruction of
network topology allows one to understand properties that might remain hidden
without the model or with a less relevant model.\\

These type of models can be enriched by adding information on nodes or edges.
For instance, `$+$' or `$-$' labels on edges may be used in order to indicate
positive or negative regulatory influences.  The existence of an edge might be
specified as being conditional on the object being studied (for instance a
cell) being in a specific global state, or on a particular gene that regulates
that particular interaction being expressed above a given threshold. These
latter types of additional information, however, refer implicitly to notions
of state and temporal evolution, and thus lead naturally towards qualitative
dynamical models.\\

Different reverse-engineering methods for topology identification differ on
the types of graphs considered. For example, in the work 
in~\cite{Fuente,Rice,Pournara,Yu,Beal,Nariai,Zou,Dojer},
edges represent statistical correlation between variables. 
In~\cite{Friedman,Jarrah,Karp,Krupa}, edges represent causal relationships 
among nodes.

\vspace*{0.2in}
\noindent\underline{\bf (b) Network Dynamical Models}

\vspace*{0.1in}
Dynamical models represent the time-varying behavior of the different
molecular components in the network, and thus provide a more accurate
representation of biological function.

Models can be used to simulate the biological system under study.  Different
choices of values for parameters correspond either to unknown system
characteristics or to environmental conditions.  The comparison of simulated
dynamics with experimental measurements helps refine the model and provide
insight on qualitative properties of behavior, such as the identification of
steady states or limit cycles, multi-stable (\EG, switch-like)
behavior, the characterization of the role of various parts of the network in
terms of signal processing (such as amplifiers, differentiators and
integrators, logic gates), and the assessment of robustness to
environmental changes or genetic perturbations.

Examples of this type of inference include those leading to various types of
Boolean networks~\cite{Liang,Akutsu,Mehra,Martin}
or systems of differential equations~\cite{Gardner,Sontag_biotech,Kim}, 
as well as multi-state discrete models~\cite{Laubenbacher}.

Depending upon the type of network analyzed, data availability and quality,
network size, and so forth, the different reverse engineering methods offer
different advantages and disadvantages relative to each other.  In Section~\ref{Benchmarking}, 
we will explore some of the common approaches to their
systematic evaluation and comparison.

\subsection{Evaluation of the Performance of Reverse Engineering Methods}
\label{metrics}

The reverse-engineering problem is by its very nature highly ``ill-posed'', in
the sense that solutions will be far from unique.  This lack of uniqueness
stems from the many sources of uncertainty: measurement error, lack of
knowledge of all the molecular species that are involved in the behavior being
analyzed (``hidden variables''), stochasticity of molecular processes, and so
forth.  In that sense, reverse-engineering methods can at best provide
approximate solutions for the network that one wishes to reconstruct, making
it very difficult to evaluate their performance through a theoretical study.
Instead, their performance is usually assessed empirically, in the following
two ways:
\begin{description}
\item[Experimental testing of predictions:] after a model has been inferred,
the newly found interactions or predictions can be tested experimentally for
network topology and network dynamics inference, respectively.

\item[Benchmarking testing:] this type of performance evaluation consists
on measuring how "close" the method of our interest is from recovering a known
network, referred to as the {\bf``gold standard"} for the problem.  
In the case of dynamical models, one evaluates the ability of the method of 
interest to reproduce observations that were not taken into account in the
``training'' phase involved in the construction of the model.  On the another 
hand, for methods that only reconstruct the network topology (wiring diagram),
a varierty of standard metrics may be applied.
\end{description}

\vspace*{0.2in}
\noindent
{\bf Metrics for Network Topology Benchmarking}

\vspace*{0.1in}
Suppose that $\Gamma$ is the graph representing the network topology  of a 
chosen ``gold standard'' network. Let $\Gamma_i$ be the graph representing
the inferred network topology. Each one of the interactions in $\Gamma_i$ can 
be classified into one of the these four classes, when comparing to the gold standard:
\begin{description}
	\item[(a)] Correct interactions inferred (true positives, TP)
	\item[(b)] Incorrect interactions inferred (false positives, FP)
	\item[(c)] Correct non-interactions inferred (true negatives, TN)
	\item[(d)] Incorrect non-interactions inferred (false negatives FN)
\end{description}

\begin{figure}[htbp]
\begin{center}
\includegraphics[width=4.5in,height=4.5in]{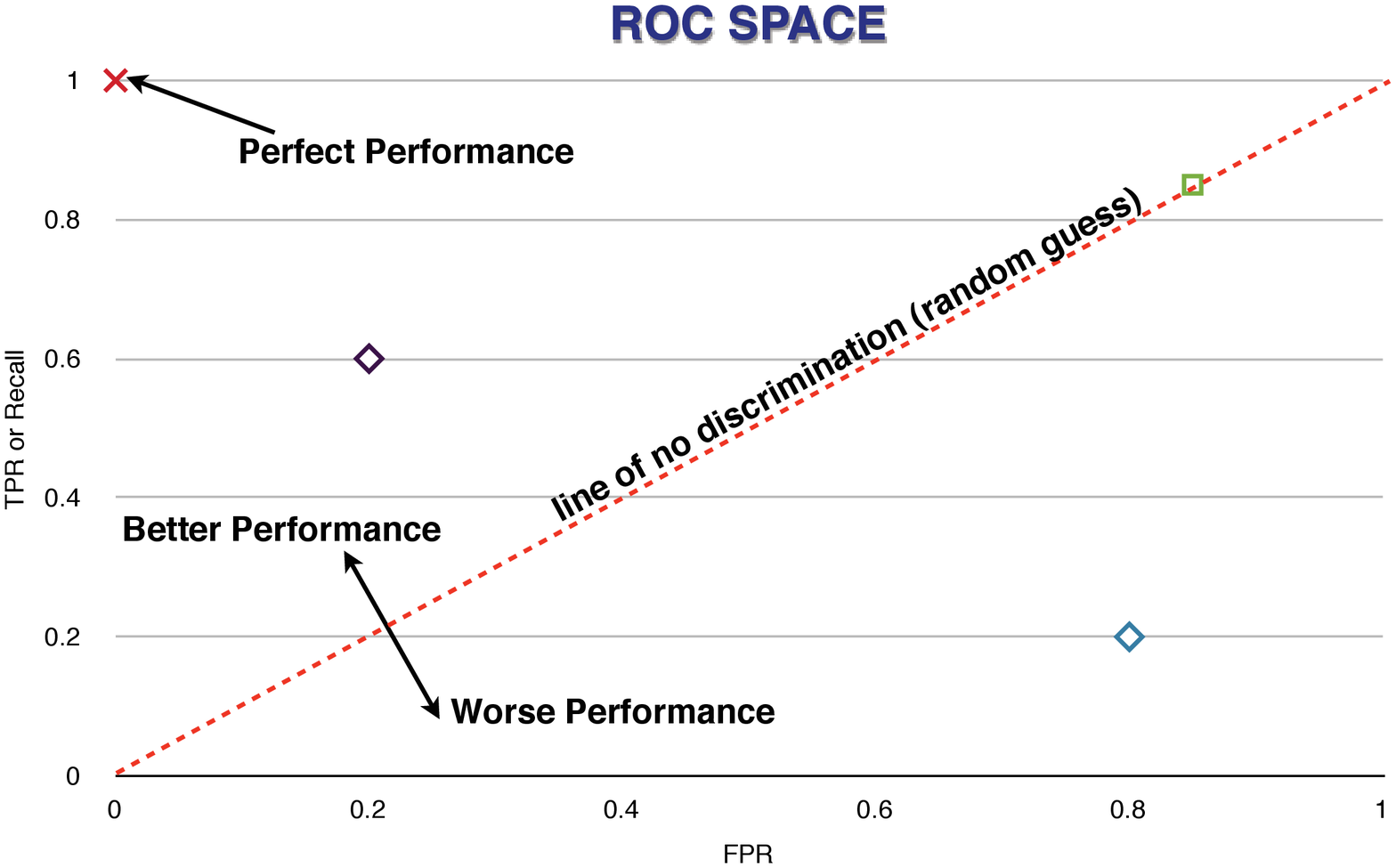}
\caption{\label{ROC} Receiver operating characteristic -ROC-space. Defined by  $FPR$ vs. $TPR$ values in a
two dimensional coordinate system: a perfect reverse engineering method will ideally have score $(\mbox{FPR},
\mbox{TPR})=(0,1)$ whereas the worst possible network will have coordinates
$(\mbox{FPR}, \mbox{TPR})=(1,0)$ and scores below the identity line (diagonal)
indicate methods that perform no better than a random guess.}
\end{center}
\end{figure}

From this classification of the interactions, we compute the following metrics:
\begin{itemize}
\item The Recall or True Positive Rate $TPR= TP / (TP + FN)$
\item The False Positive Rate $FPR = FP / (FP + TN)$.
\item The Accuracy $ACC = (TP + TN) / TotI$ where $TotI$ is the total number
  of possible interactions in a network.
\item The Precision or Positive Predictive Value $PPV = TP/(TP+FP)$. 
\end{itemize}
As mentioned earlier, the reverse-engineering problem is underconstrained.
Every algorithm will have one or more free parameters that helps select a
``best" possible prediction. 
Hence, a more objective evaluation of performance has to somehow involve
a range of parameter values. 
One way to evaluate performance across ranges of parameters is the
{\bf receiver operating characteristic (ROC)} method, based on
the plot of $FPR$ vs. $TPR$ values.
The resulting {\bf ROC plot} depicts relative 
trade-offs between true positive predictions and false positive prediction across different parameter 
values (See Fig.~\ref{ROC}). 
A closely related approach is the
{\bf Recall-Precision plot}, obtained by plotting $TPR$ vs. $PPV$ values. 

\section{Classical Combinatorial Algorithms: A Case Study}
\label{methods}

We have briefly discussed some basic aspects of reverse-engineering of
biological systems.  Next, as a case of study, we focus our attention on some
reverse-engineering algorithms that rely upon the solution of the so-called
``Hitting Set Problem".  The Hitting Set Problem is a classical problem in
combinatorics and computer science.  It is defined as follows:

\begin{problem}[HITTING SET Problem] 
Given a collection $\mathscr{H}$ of subsets of $E=\{1,\ldots,n\}$,
find the smallest set $L\subseteq E$ such that $L \cap\mathscr{X} \neq \emptyset$ for all $\mathscr{X} \in \mathscr{H}$.
\end{problem}

The Hitting Set problem is NP-hard, as can be shown via transformation from its
dual, the (Minimum) Set Cover problem~\cite{Karp2}.

We next introduce some reverse engineering methods based on the hitting set
approach. 
\begin{itemize}
\item Ideker \EA~\cite{Karp}. 

This paper introduces two methods to infer the
topology of a gene regulatory network from gene expression measurements.  The first
``network inference'' step consists of the estimation of a set of Boolean
networks consistent with an observed set of steady-state gene expression
profiles, each generated from a different perturbation to the genetic
network studied.  Next, an ``optimization step'' involves the use of an
entropy-based approach to select an additional perturbation experiment in
order to perform a model selection from the set of predicted Boolean
networks. In order to compute the sparsest network that interpolates the data,
Ideker \EA rely upon the
``Minimum Set Cover'' problem.  
An approximate solution for the Hitting Set problem is obtained by
means of a branch and bound technique~\cite{Nem}. Assessment is performed
``\emph{in Numero}'': the proposed method is evaluated on simulated
networks with varying number of genes and numbers of interactions per gene.

\item Jarrah \textit{et al.}~\cite{Jarrah} 

This paper introduces a method for the inference of the network topology from gene
expression data, from which one extracts state transition measurements of
wild-type and perturbation data.  
The goal of this reverse-engineering algorithm is to output one or more most likely network topologies 
for a collection $x_1, \dots , x_n$ of molecular species (genes, proteins, etc), 
which we will refer to as variables.  The state of a molecular species can represent its 
levels of activation.  That is, each variable $x_i$ takes values in the set $X = \{0, 1, 
2,\dots\}$    and the interactions among species indicate causal relationships among
molecular species. The inference algorithm 
takes as input one or more time courses of observational data.  The output is a most likely 
network structure for the interactions among $x_1,  \dots , x_n$ that is consistent with the observational data:  The notion 
of consistency with observational data makes the assumption that the
regulatory network for
$x_1,  \dots, x_n$ can be viewed as a dynamical system that is 
described by a function $f: X_n \rightarrow X_n$, which transforms an input state $(s_1, \dots , s_n)$, $s_i \in X$, of 
the network into an output state $(t_1,  \dots , t_n)$ at the next time step.  A directed edge $x_i \rightarrow x_j$ in 
the graph of the network topology of this dynamical system $f$ indicates that the value of $x_j$ under 
application of $f$ depends on the value of $x_i$.  Hence a directed graph is 
consistent with a given time course $s_1,  \dots , s_r$ of states in $X_n$, if it is the network topology of 
a function $f: X_n \rightarrow X_n$ that reproduces the time course, that is, $f(s_i) = s_{i+1}$ for all $i$. 

One possible drawback of reverse engineering approaches lies in the fact that
they construct the ``sparest'' possible network consistent with the given data.
However, real biological networks are known to be not minimal~\cite{TSE99}. 
Although accurate measures of deviation from sparsity are difficult to estimate,
nonetheless it seems reasonable to allow additional edges in the network in a 
``controlled'' manner that is consistent with the given data.
As already commented in~\cite{Karp}, it is possible to add redundancies to the
reverse engineering construction. The basic hitting set approach provides only
a minimal set of connections, whereas real biological networks are known to
contain redundancies (\EG, see~\cite{TSE99}).  To account for this, one can
modify the hitting set approach to add redundancies systematically by allowing
additional parameters to control the extra connections. Theoretically, in terms
of the algorithm this
corresponds to a standard generalization of the set-cover problem, known as
the set-multicover problem, which is well-studied in the literature, and for
which approximation algorithms are known~\cite{BDS04}.

The search for the topologies that interpolate the input data involves directly the Hitting Set problem, 
which is solved analytically with the use of a computational algebra tools.
\end{itemize}
The algorithms presented in~\cite{Krupa,Sontag2} also make use
of Hitting Set algorithms, but we will restrict our attention to the
comparison of the two methods described above.

\subsection{Benchmarking RE Combinatorial-Based Methods}
\label{Benchmarking}

\subsubsection{\textit{In Silico} Gene Regulatory Networks}\label{insil}

We use data from two different regulatory networks.  These contain some
features that are common in real regulatory networks, such as time delays and
the need for a measurement data presented
into discrete states ($0,1,2, \dots $).

\medskip

\noindent
{\bf In Silico Network 1: Gene Regulatory Network with External Perturbations}.
This network was originally introduced in~\cite{camacho}. It was 
generated using the software package given in~\cite{mendes03},
the interactions between genes in this regulatory network are
phenomenological, and represent the net effect of transcription,
translation, and post-translation modifications on the regulation of the genes
in the network.  The model is implemented as a system of ODEs in
\textit{Copasi} \cite{hoops}.
\medskip

\begin{figure}[htbp]
\begin{center}
\includegraphics[width=3.5in]{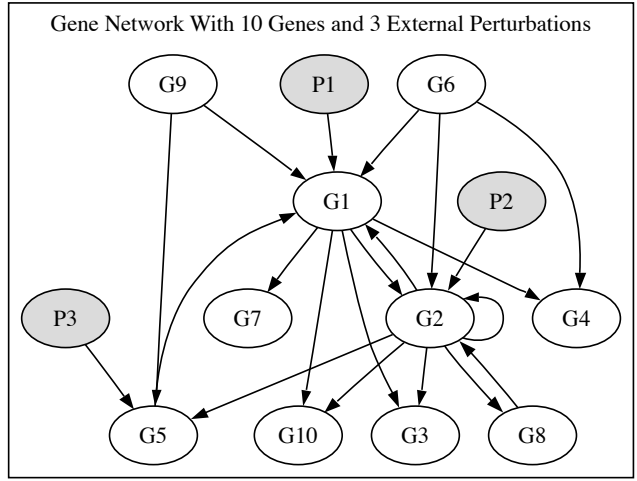}
\caption{\label{insilico1} Network 1:  $10$ genes and $3$ environmental perturbations. 
In this network, the $3$ environmental perturbations P1, P2 and P3  directly affect the expression rate of genes 
G1, G2 and G5, respectively.} 
\end{center}
\end{figure}

This network, shown in Fig.~\ref{insilico1}, consists of $13$ species: ten genes
plus three different environmental perturbations. The perturbations affect the
transcription rate of the gene on which they act directly (through inhibition
or activation) and their effect is propagated throughout the network by the
interactions between the genes. 

\vspace*{0.2in}
\noindent
{\bf Network 2: Segment Polarity Genes Network in \textit{D. melanogaster}}.
The network of segment polarity genes is responsible for
pattern formation in the \textit{Drosophila melanogaster} embryo. 
Albert and Othmer~\cite{Albert} proposed and analyzed a Boolean model based on
the binary ON/OFF representation of mRNA and protein levels of five segment
polarity genes. This model was constructed based on the known topology and it
was validated using published gene and expression data.  We 
generated time courses from this model, from which we will attempt to
reverse-engineer the network in order to benchmark the performance of the reverse-engineering 
algorithms being evaluated.

The network of the segment polarity genes represents the last step in the
hierarchical cascade of gene families initiating the segmented body of the
fruit fly. The genes of this network include engrailed ($en$), wingless
($wg$), hedgehog ($hh$), patched ($ptc$), cubitus interruptus ($ci$) and
sloppy paired ($slp$), coding for the corresponding proteins, which are
represented by capital letters ($EN, WG, HH, PTC, CI$ and $SLP$).  Two
additional proteins, resulting from transformations of the protein $CI$, also
play important roles: $CI$ may be converted into a transcriptional activator,
$CIA$, or may be cleaved to form a transcriptional repressor $CIR$. The
expression of the segment polarity genes occurs in stripes that encircle the
embryo.  These key features of these patterns can be represented in one
dimension by a line of $12$ interconnected cells, grouped into $3$ parasegment
primordia, in which the genes are expressed every fourth cell. In Albert and
Othmer~\cite{Albert}, parasegments are assumed to be identical, and thus only
one parasegment of four cells is considered. Therefore, in the model, the
variables are the expression levels of the segment polarity genes and proteins
(listed above) in each of the four cells, and the network can be seen as a
$15\times 4 = 60$ node network.  Using the wild-type pattern
from~\cite{Albert}, we consider one wild-type time series of length $23$.
	
\begin{figure}[htbp]
\begin{center}
\includegraphics[width=3.5in]{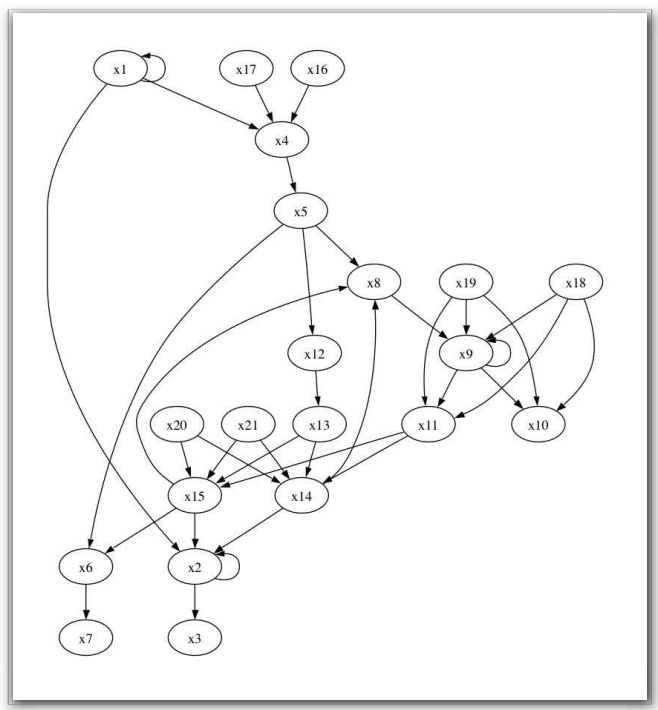}
\caption{Segment Polarity Genes Network on the {\textit D. melanogaster}. This
network consists of the interaction of $60$ molecular species: genes and
proteins.}  
\label{FruitFly}
\end{center}
\end{figure}

\subsubsection{Results of Comparison}

\begin{table}[htbp]
\begin{center}
\begin{tabular}{|lcccccccc|}
\hline
                                        & TP    & FP    & TN    & FN    & TPR   & FPR   & ACC   & PPV\\
\hline
Network1 &&&&&&&&\\
Jarrah Exact Sol D7             &12             &34             &113    &10             &.5454  &.231   &.7396  &.2608\\
Jarrah Exact Sol Q5             &9              &49             &98             &13             &.4090  &.3334  &.6331  &.1551\\
Jarrah Exat Sol I5              &7              &46             &101    &15             &.3181  &.313   &.6390  &.1320\\
Karp Greedy Approx R1   &9              &49             &98             &13             &.4090  &.666   &.016   &.084\\
Karp Greedy Approx R2   &11             &63             &84             &11             &.5             &.571   &.020   &.115\\
Karp LP Approx. R1              &7              &46             &101    &15             &.318   &.687   &.014   &.064\\
Karp LP Approx. R2              &9              &59             &88             &13             &.409   &.598   &.018   &.092\\
\hline
Network 2 &&&&&&&&\\
Jarrah Exact Sol                & --            &--             &--             &--             &--             &--             &--             &--\\
Karp Greedy Approx       R1     &4              &3321   &91             &124    &.031   &.026   &.923   &.042\\
Karp Greedy Approx R2   &15             &3254   &218    &113    &.117   &.062   &.908   &.064\\
Karp LP Approx. R1              &3              &3279   &93             &125    &.023   &.026   &.939   &.031\\
Karp LP Approx. R2              &9              &3285   &187    &119    &.070   &054    &.915   &.045\\
\hline
\end{tabular}
\caption{\textbf{Comparison of RE methods}}
\label{results}
\end{center}
\end{table}

In this section we compare the results obtained after running Jarrah~\EA's
and Ideker\EA's methods on each of the above networks. Computations
were made on Mac OS X, Processor  2GHz Intel Core 2 Duo.

As we mentioned in Section~\ref{methods}, for Jarrah \EA's method, the input data must be
discrete. Hence in order to apply this reverse-engineering method to network 1 we discretize the input data,
considering then different discretizations as our running parameter to test Jarrah\EA's method in
the ROC space. We specifically use three discretization methods: a graph-theoretic based approached ``D''
(see~\cite{Polynome}), as well as quantile ``Q'' (discretization method on which each variable state receives
an equal number of data values) and interval ``I'' discretization (discretization method on which we select
thresholds for the different discrete values).

For Ideker \EA's  method we have considered both Greedy and 
Linear Programming approximations to the Hitting set problem as well as 
redundancy values (how many extra edges one allows) of $R=1$ or $2$.

We have displayed some our results on Table~\ref{results}. We observe that for network 1,
Jarrah \emph{et.~al.}'s method obtains better results than Ideker
\emph{et.~al.}'s method when 
considering these values in the ROC space,  although both fare very poorly.  On the another hand, we observe that Ideker \emph{et.~al.}'s
method achieves a performance no better than random guessing on this network. In contrast,
for network 2, Jarrah \EA's method could not obtain any results after running their
method for over 12 hours, but Ideker \EA's method was able to compute results for such
network in less than 1 minute. Also Ideker \EA's method improved slightly its results when
the redundancy number is increased; this might
indicate the shortcoming of inferring sparser networks when they are of larger
size containing redundancies.

\subsection{Software Availability}

The implementation of Jarrah \emph{et.~al.}'s algorithm~\cite{Jarrah}
is available online through the web interface provided at
\url{http://polymath.vbi.vt.edu/polynome/}. The implementation of
Ideker \EA's algorithm~\cite{Karp} is available online through
the web interface provided at \url{http://sts.bioengr.uic.edu/causal/}.

\section{Concluding Remarks}

In this chapter, we first provided a brief discussion of the biological
reverse-engineering problem, which is a central problem in systems biology. As
a case study, we then focused on two methods that rely upon the solution of
the ``Hitting Set problem'', but which differ in their approach to solve this
problem, thus leading to different performance.

In terms of network inference power, we hypothesize that, for the smaller
network, the poor quality of the results when using Jarrah's approach might be
ascribed to the type of data used: in~\cite{Jarrah} it is claimed that the
method performs better if perturbation data is added.  The algorithm has the
ability of considering both wild-type and mutant data to infer the network,
and probably results would improve if using such additional data.  In the case
of Ideker \emph{et.~al.}'s method, in both networks we think that it is possible that the low
quality of results could be due to lack of ability of using more than one time
series at a time, as well as the fact that the implementation of the method
does not include self loops
(self-loops are edges connecting a node to itself which may, for example, 
represent degradation terms in biochemical systems). We believe that this feature is fundamental for a good
performance of the algorithm.

When comparing the computational efficiency of the approaches, one should keep
in mind that there will always be a difference between exact solutions and
approximate solutions based upon greedy algorithms or linear programming
relaxations.  
However, since the size of the networks was fairly small, it is possible that
the reason for which Jarrah's method did not find a solution within a
reasonable time might lie in encoding issues rather than intrinsic
computational complexity of the problem.  

\paragraph{Acknowledgments}
The authors would like to thank Joe Dundas for the implementation and
maintenance of the web tool for Ideker \emph{et.~al.} method. We would like to thank as well
Dr. Brandilyn Stigler for useful discussions on different aspects of this book
chapter.  
This work was supported in part by grants
AFOSR FA9550-08,
NIH 1R01GM086881,
and NSF grants
DMS-0614371,
DBI-0543365, IIS-0612044,
IIS-0346973 and the DIMACS special focus on Computational and Mathematical
Epidemiology.

\end{document}